\documentclass[aps,rmp,reprint,amsmath,amssymb,graphicx,shortbibliography]{revtex4-1}
\usepackage[utf8]{inputenc}
\usepackage{bm}
\usepackage{amsmath}
\usepackage{graphicx}
\usepackage{tabularx}
\usepackage{float}
\usepackage{epstopdf}
\usepackage{natbib}
\usepackage{array}
\usepackage{braket}
\usepackage[usenames,dvipsnames]{color}
\usepackage{dcolumn}
\usepackage{bm}
\usepackage{soul}  
\usepackage{changes}
\usepackage{hyperref}
\usepackage{enumitem}
\usepackage{booktabs}
\hypersetup{colorlinks=true, linkcolor=blue, filecolor=blue, urlcolor=blue}
\usepackage{xcolor}
\definecolor{darkblue}{RGB}{0, 0, 139}
\definecolor{darkgreen}{RGB}{0, 100, 0}
\definecolor{darkred}{RGB}{139, 0, 0}
\usepackage{iftex}

\ifTUTeX
  \usepackage{fontspec}
\else
  \usepackage[T1]{fontenc}
  \usepackage[utf8]{inputenc} 
  \DeclareUnicodeCharacter{200B}{{\hskip 0pt}}
\fi

\begin{document}
	
\title{All-Photonic and Memory-Based Quantum Repeaters Enabled by Quantum Dots}
%
\author{Joanna M. Zajac$^{1,2}$}
\affiliation{\mbox{$^{1}$Current: Independent Researcher, United Kingdom}}
\affiliation{\mbox{$^{2}$Former affiliation: Brookhaven National Laboratory, Upton, NY 11973, USA}}
\author{Reza Hekmati, Tobias Huber-Loyola, Sven H\"{o}fling}
\affiliation{\mbox{Julius-Maximilians-University of Würzburg, Lehrstuhl für Technische Physik, Germany}}
	
\date{\today{}}
	
\begin{abstract}
This review surveys recent progress in III--V semiconductor quantum dots\,(QDs) as a platform for quantum repeaters. We start by 
discussing the state of the art in QD-based non-classical light sources. 
Specifically, we report on on single-photon and entangled-pair sources operating across near-
infrared and telecom wavelengths, with emphasis on the key metrics---multi-photon suppression \(g^{(2)}(0)\), photon 
indistinguishability, extraction efficiency, brightness, and spin coherence time---while discussing frequency conversion, excitation schemes, cavity engineering, remote indistinguishability, and spin coherence.

We then examine the two principal repeater architectures. For all-photonic repeaters we review linear cluster- and graph-state generation 
using QD spins, recent experimental milestones, and the critical role of spin dephasing time. For memory-based repeaters we focus on 
heterogeneous implementations combining deterministic QD photon sources with room-temperature alkali-vapor memories, providing rate 
benchmarking against other platforms, discussion of storage protocols, wavelength compatibility, and early demonstrations. 
Enabling technologies such as cryogenic cooling, on-chip photonic integration, network synchronization and multiplexing are 
also presented.

The review highlights the strength of QD-based architectures and identifies the remaining milestones required for their deployment in practical fiber-based quantum networks.
\end{abstract}
	
\maketitle
\tableofcontents{}

\section{Introduction Quantum Repeaters}
\label{sec:QuantRepOverview}
The realization of a quantum internet relies on distributing entanglement across large-scale quantum networks \cite{KimbleNature08,RitterRempeNature12,HansonWehnerScience18}. Semiconductor quantum dots have emerged as a leading platform in this context, owing to their potential for deterministic, high-rate single-photon generation \cite{LuPanNatNano21,trottaMatQuantTech21,MichlerWaksLiuNatNano23,HeindelAdvQuantTech22,LoockAdvQuantum20}.\\
A fundamental obstacle to long-distance operation is photon loss in optical fibers. In the O- and S-bands, attenuation still causes exponential suppression of transmission rates, reducing even GHz photon streams to negligible levels over just a couple hundred kilometers. This limitation cannot be overcome by direct amplification: photon loss is irreversible, and any attempt to amplify unknown quantum states is forbidden by the no-cloning theorem \cite{Park70,WootterZurekNature82}.\\
To overcome these limitations, the quantum repeater protocol was introduced by Briegel, D\"{u}r, Cirac, and Zoller (BDCZ protocol)     \cite{BriegelZollerPRL98}. This seminal work, along with subsequent extensions using improved schemes     \cite{DuanLukinCiracZollerNature01, SangouardGisinPRA07, SangouardGisinRevModPhys11}, established the foundation for long-distance quantum communication.
Quantum repeater networks (QRNs) enable quantum communication over long distances by dividing the total channel length $L_{\rm tot}$ into $N=2^n$ shorter elementary links of length $L_0 = L_{\rm tot}/N$, where $n$ denotes the nesting level. Entanglement is first generated between adjacent nodes and then extended across the full distance through successive entanglement swapping operations.
Crucial to the operation of multi-node repeater chains is the use of quantum memories that can store entanglement until successful swapping is heralded on each segment. Once confirmed, the stored states are read out and swapped to establish entanglement over longer distances. This architecture therefore requires heralded entanglement generation (e.g., via protocols such as Cabrillo \textit{et al.} \cite{CabrilloZollerPRA99}) as well as entanglement purification to mitigate accumulated decoherence.

The primary goal of current quantum repeater research is to outperform direct transmission. Twin-field quantum key distribution (TF-QKD) \cite{LucamariniNature18} can surpass the direct-transmission rate limit over long distances without quantum memories, with demonstrations reaching up to 1000\,km \cite{ChenNatPhot21,WangNatPhot22,LiuPRL23}. However, TF-QKD does not provide a general architecture for distributing entanglement across arbitrary multi-node quantum networks. Quantum repeaters therefore remain necessary for scalable entanglement distribution and for applications beyond point-to-point key generation.

The achievable rates vary strongly between repeater protocols and physical platforms. In this review, we focus on the potential of heterogeneous architectures that combine quantum-dot photon sources, which offer deterministic and high-rate emission, with room-temperature alkali-vapor quantum memories, which provide a technically attractive memory platform with relatively simple operation. Such hybrid systems are promising for long-distance quantum networks, but their practical use requires overcoming key interface challenges, in particular bandwidth matching, wavelength compatibility, noise suppression, and efficient storage of single photons.

\section{Quantum Dot Photon Sources for Quantum Repeaters}
\label{sec:QDs}

Recent advances in III--V semiconductor quantum dots have established them as one of the most mature solid-state platforms for deterministic quantum light generation. State-of-the-art devices now approach the ideal single-photon emitter, with near-unity indistinguishability under resonant excitation, very low multi-photon contribution, and high extraction efficiencies enabled by cavity quantum electrodynamics~\cite{SenellartNatureNano17, HuberLoyolaAPR2026}. In parallel, significant progress has been made toward entangled photon-pair generation and the extension of emission wavelengths into the telecom bands, which is essential for fiber-based quantum networks~\cite{MichlerWaksLiuNatNano23, HuberLoyolaAPR2026}.

Despite these advances, key challenges remain, including the simultaneous optimization of brightness and coherence, performance degradation at telecom wavelengths, and integration into scalable architectures compatible with quantum repeater protocols. 

For quantum repeater applications, the most relevant figures of merit are  brightness, multi-photon emission probability, photon indistinguishability, on-demand operation, spin coherence time, and entanglement fidelity. High brightness and repetition rate directly determine the entanglement generation rate, while low multi-photon emission (quantified by $(g^{(2)}(0)$) suppresses errors in heralded protocols. Near-unity photon indistinguishability is critical for high-visibility two-photon interference in Bell-state measurements.
Extending $(T_2)$ through resonant excitation, dynamical decoupling, or nuclear-spin control is therefore as critical as improving brightness and lowering \(g^{(2)}(0)\).

The remainder of this section is organized as follows. We first review near-infrared (NIR) quantum dots in Sec.~\ref{sources}, 
covering single-photon sources, excitation schemes, and single-photon frequency conversion. We then discuss telecom-wavelength 
emitters in Sec.~\ref{TelecomQDs}. Finally, Sec.~\ref{Protocols} addresses key building blocks for quantum repeater operation, 
including the extension of spin coherence times, achieving high indistinguishability between remote quantum dots, and the 
realization of spin-photon gates for entanglement generation and teleportation.
\subsection{Photon sources at Near-Infrared}
\label{sources}
Semiconductor quantum dots for quantum repeater applications are predominantly grown by molecular beam epitaxy (MBE) or metal organic vapor phase epitaxy (MOVPE). The Stranski--Krastanov mode in the InGaAs/GaAs system, driven by strain relaxation due to the $\sim$7\,\% lattice mismatch, remains the most widely used and established growth technique for near-infrared emission~\cite{michler2009single}.
This process leads to the formation of islands that are typically anisotropic, elongated along $\langle 110 \rangle$ and $\langle 1\overline{1}0 \rangle$ directions. In contrast, droplet epitaxy~\cite{LeeKoguchiJJAplPhys98} and droplet etching~\cite{RastelliPhysE04} enable the formation of highly symmetric GaAs/AlGaAs dots with strongly reduced fine-structure splitting.  Considerable effort has also been directed toward deterministic positioning of individual QDs via buried stressors~\cite{LimameAPL2023}, pre-patterned substrates~\cite{KiravittayaSchmidtPL06,SchneiderNanotech08,JuskaNaturePhoton2013}, Au-seeded nanowires~\cite{DalacuNanotech2009,Heinrich10}, and selective-area growth~\cite{OzakiASS2008}, all of which are essential for scalable integration into photonic circuits required for practical repeater architectures.
Strain engineering, including strain-reducing layers\cite{FioreLangbeinIlegemsAPL2000} and metamorphic buffers\cite{SeravaliiFrigeriCrystalResTech14,WronskiMat21}, is routinely employed to red-shift the emission wavelengths and will be discussed in Sec.\ref{TelecomQDs}.
    
\textbf{Cavity-enhanced designs} dominate high-performance quantum dot single-photon sources because they dramatically improve photon 
extraction efficiency and allow tuning of the emission lifetime via the Purcell effect. Key architectures include narrow-linewidth 
micropillars, open cavities and photonic crystal cavities, as well as broadband designs such as solid-immersion lenses, circular Bragg 
gratings (also known as bullseye cavities), and tapered micro-pillars, for review see Ref.\cite{SenellartNatureNano17, HuberLoyolaAPR2026}.
In the weak-coupling regime, these cavities provide Purcell enhancement of the spontaneous emission rate according to $F_P = \frac{3}
{4\pi^2} (\lambda/n)^3 (Q/V)$, where $\lambda$ is the operating wavelength, $n$ the refractive index, $Q$ the quality factor, and $V$ the 
mode volume. 
A wide range of photonic architectures have been developed to enhance photon extraction from NIR In(Ga)As/GaAs quantum dots, each reflecting 
different trade-offs between efficiency, spectral selectivity, and scalability. Micropillar cavities remain among the most mature cavity-QED 
platforms, combining Purcell enhancement with high extraction efficiencies approaching 70\%, while simultaneously enabling near-unity 
indistinguishability under resonant excitation \cite{DingPRL16,GinesPRL22}. Circular Bragg gratings offer a complementary broadband approach 
with relaxed spectral alignment requirements and have demonstrated polarized single-photon collection efficiencies exceeding 50\% \cite{WangNatPhoton2019,PeniakovLaser2024}. Solid immersion lens concepts improve free-space extraction without relying on narrow cavity resonances, reaching brightness values of 40--50\% for moderate numerical apertures \cite{FischbachACS17,AhnAPL23,MaZajacOptLett15}, while optimized geometries can push these values significantly further. Photonic crystal waveguides represent a particularly attractive route for integrated architectures, achieving near-unity emitter-to-waveguide coupling efficiencies ($\beta \approx 98\%$) and enabling deterministic routing of single photons on-chip \cite{ArcariPRL14}. Finally, open cavity platforms provide a highly flexible alternative, allowing in-situ spectral and spatial tuning of the cavity-emitter coupling, which is particularly attractive for overcoming fabrication variability and exploring coherent light-matter interfaces with record extraction efficiencies exceeding $90\%$\cite{TommWarburtonNatNano2021,DingNatPhoton2025}, albeit currently with less mature scalability compared to monolithic nanophotonic devices. Taken together, these results illustrate the substantial progress achieved in photonic engineering of single quantum dot emitters under cryogenic, resonant, or pulsed excitation conditions, with photon extraction efficiencies now approaching levels required for practical high-performance quantum photonic applications.

\textbf{Electrically contacted devices} are highly desirable for practical single-photon sources because they enable charge control, Stark tuning, and, in principle, direct electrical injection compatible with on-chip integration. Fully electrically driven quantum-dot sources have demonstrated attractive operating rates, but their photon quality still lags behind optically resonant sources. Schlehahn \textit{et al.} reported 61\% extraction efficiency at 625\,MHz repetition rate, with $g^{(2)}(0)$ between 0.076 and 0.2 for repetition rates up to 1.2\,GHz and indistinguishability of 0.4 at $(92\pm23)\,\mathrm{Mcps}$\cite{SchlehahnAPL16}. Even higher electrical drive rates reaching 2\,GHz repetition rate, with $g^{(2)}(0)\approx 0.27$, were reported by Hargart \textit{et al.} in InP/GaInP quantum-dot diodes\cite{HargartAPL13}. These values illustrate the scalability potential of electrical injection, but also its current limitations: carrier capture and relaxation processes introduce timing jitter and excess charge noise, which degrade multi-photon suppression and photon indistinguishability.

For the highest-performance devices, electrical contacts are therefore often used not for carrier injection but for charge stabilization and frequency tuning via the quantum-confined Stark effect, while the quantum dot is still driven optically and resonantly. This operating mode combines the coherence advantages of resonant excitation with the stability of diode structures. For example, resonantly driven charge-tunable devices have achieved $g^{(2)}(0)=0.019$ with no blinking on long timescales\cite{ZhaiNatComm20}.

\textbf{The excitation scheme} fundamentally determines the performance of quantum dot single-photon sources and therefore their suitability for quantum networking applications. In particular, the choice of excitation directly impacts multi-photon contributions, indistinguishability, and usable source brightness, which together define the performance of interference-based protocols such as entanglement swapping in quantum repeaters. A clear hierarchy has emerged across different excitation approaches. 
Non-resonant excitation, while experimentally straightforward, introduces carrier relaxation processes that lead to timing jitter and additional dephasing, substantially degrading photon coherence. This is reflected in comparatively high multi-photon contribution, with measured values such as $g^{(2)}(0)=0.26$ under above-band excitation, while semi-resonant p-shell excitation partially mitigates these effects and improves the multi-photon contribution to $g^{(2)}(0)=0.077$ \cite{LeeWaksNanoLett20}.

The highest performance is consistently achieved when the quantum dot is driven resonantly. By directly addressing the optical transition, resonant excitation avoids incoherent relaxation pathways and enables near-ideal photon generation. Landmark demonstrations of cavity-enhanced resonant excitation have reported ultra-low multi-photon emission probabilities ($g^{(2)}(0)=0.0028$), near-unity indistinguishability (0.9956), and simultaneously high extraction efficiencies approaching 66\% \cite{SomaschiNatPhoton16,DingHoflingPanPRL16}. Resonant fluorescence experiments have further been instrumental in revealing the intrinsic coherence properties of quantum dots, demonstrating transform-limited linewidths and clarifying the role of charge noise, nuclear spin fluctuations, and phonon interactions \cite{MatthiesenPRL12,KuhlmannWarburtonNatPhys13,MaleinPRL16}. More recently, cavity-enhanced resonance fluorescence has shown that these excellent coherence properties can be retained while improving collection efficiency \cite{RickertACS25,TommWarburtonNatNano2021}.
  
At the same time, strictly resonant excitation introduces practical challenges, most notably the need to suppress scattered laser light, which often requires polarization filtering and can reduce the usable brightness. This has motivated the development of more robust coherent excitation schemes that seek to preserve the advantages of resonant driving while relaxing some of its experimental constraints. Adiabatic rapid passage using chirped pulses, for example, offers a more stable population inversion against pulse-area fluctuations while still producing highly indistinguishable photons, with reported Hong-Ou-Mandel visibilities approaching 98\% \cite{WeiHoflingPanNanoLett14}. Similarly, phonon-assisted excitation exploits controlled coupling to the solid-state environment to achieve efficient state preparation with improved robustness \cite{QuilterFoxPRL15}. The recently proposed SUPER protocol extends this idea further, using frequency-modulated off-resonant driving to suppress re-excitation while maintaining high coherence \cite{BrachtReiterPRX21,KarliNanoLetter2022,BoosAQT24}. Raman-based approaches provide yet another route to highly coherent photon generation, with indistinguishabilities up to 0.98, although historically at the expense of reduced emission rates \cite{HeHoflingPanPRL13}. 

It is worth noting that even under resonant excitation, photon indistinguishability can be limited by the coherence of the driving laser when operated in continuous-wave mode, as characterized by the coalescence time window \cite{ProuxDiederichsPRL15}. This underscores that both the emitter properties and the choice of excitation conditions must be optimized together to achieve high interference visibility.

Taken together, these developments illustrate that excitation engineering has become a central optimization lever for quantum dot photon sources. While resonant excitation currently defines the state of the art in photon quality, the broader challenge is to identify schemes that retain this level of coherence while simultaneously delivering high brightness, operational robustness, and compatibility with scalable photonic architectures \cite{JoosPortalupiMichlerNanolett24}.
\\ \textbf{Frequency Conversion into Telecom}
Quantum frequency conversion (QFC) remains the primary approach for interfacing state-of-the-art near-infrared quantum dot single-photon sources with low-loss telecom fiber networks. QFC is typically realized via difference-frequency generation in periodically-poled lithium niobate (PPLN) waveguides.\\
Recent experiments have demonstrated substantial improvements in conversion performance. Using PPLN waveguides, end-to-end conversion efficiencies reaching {48.4\%} have been achieved for the conversion of single photons from a quantum dot emitting at 925.7\,nm to the telecom C-band (1560\,nm) \cite{YouWeiPanAdvPhoton22}. A fiber-coupled system further reported an end-to-end efficiency of {40\%} while maintaining low multi-photon emission with $g^{(2)}(0) = 0.024$ and high indistinguishability $V = 94.8\%$ \cite{DaLioLodahlMidoloAdvQuantTech22}.
QFC has been shown to preserve photon indistinguishability \cite{KambsMichlerBecherOptEx16} as well as light-matter entanglement \cite{YamamotoDeGraveNature12}. Moreover, it effectively erases spectral mismatches between independent quantum dot sources caused by inhomogeneous broadening or spectral wandering and can thus be used to interfere sources that were not idetnical before QFC\cite{StrobelNatCom25}.
%
\begin{table*}[t!]
\centering
\label{tab:telecom-nir-performance}
\begin{tabular}{|l| l| c| c| l|}
\toprule
\hline
\textbf{Reference} & \textbf{Wavelength} & \(\boldsymbol{g^{(2)}(0)}\) & \textbf{Indistinguishability (HOM)} & \textbf{Operating Conditions} \\
\hline
\midrule
\cite{SomaschiNatPhoton16} & NIR (925\,nm) & 0.0028 & 0.9956 & DBR micropillar(7--8),\,resonant\,($\sim$80\,MHz,\,15\,ps) \\ 
\cite{DingHoflingPanPRL16} & NIR (898\,nm) & 0.009 & 0.985 & DBR micropillar(4--5),\,resonant\,($\sim$80\,MHz,\,3\,ps) \\ 
\cite{ZhaiWarburtonNatNano22} & NIR (GaAs) & 0.01 & 0.93 (remote) & None,\,resonant\,($\sim$80\,MHz,\,6\,ps)\\ %
\cite{WangNatPhoton2019} & NIR (890\,nm) & 0.025 & 0.975 & Micropillar,\,resonant\,(76\,MHz)\\    
\midrule
\cite{GeNanoLett24} & C-band & 0.078 & -- & Elliptical Bragg grating(EBG),\,resonant\,(76\,MHz) \\
\cite{HolewaSemenovaNatComm24} & C-band & 0.003 & 0.193 (raw) & Circular Bragg grating,\,phonon-assisted \\
\cite{HauserNatComm26} & C-band & 0.017 & 0.917 (raw) & Circular Bragg grating,\,phonon-assisted \\
\cite{BehrendsArxiv2026} & C-band & 0.007 & 0.9 (raw) & Circular Bragg grating,\,resonant\,(100\,MHz, 4\,ps) \\
\cite{KimWaksOptica16} & O-band & 0.085 & 0.18 (0.67 post-selected) & Photonic crystal(4),\,above-band \\
\bottomrule
\hline
\end{tabular}
\caption{Best reported single-photon metrics ($g^{(2)}(0)$ and indistinguishability) for NIR and telecom III--V QDs. Operating conditions include cavity type, Purcell factor, repetition rate, pulse width, and excitation scheme where available (more details and extraction efficiencies discussed in the text).}
\end{table*}

\subsection{Photon Sources at Telecom Wavelengths}
\label{TelecomQDs}
Operation in the telecom O- and C-bands is essential for fiber-based quantum communication. Two primary III-V material systems are actively pursued: InAs/InP and InAs/GaAs quantum dots (QDs) to produce photons directly in the telecom bands without the need for QFC.
   
InAs/InP QDs benefit from a moderate lattice mismatch, which enables direct growth of emitters in the telecom O- and C-bands without the need for strain-engineering. Recent progress has therefore focused on combining this material platform with nanophotonic structures that improve extraction efficiency. In quaternary InAlGaAs membranes, circular Bragg grating cavities have demonstrated Purcell factors of \(\sim 10\), together with low multi-photon emission \(g^{(2)}(0)=0.007\) and raw HOM visibility of 90\%\cite{BehrendsArxiv2026,HauserNatComm26}. Elliptical Bragg grating cavities on InP provide a complementary route toward polarized C-band emission, with Purcell enhancement of 5.25, \(g^{(2)}(0)=0.078\), and a collection efficiency of 24\% into NA=0.65, while showing no blinking\cite{HuoNanoLett24}. 

In parallel, droplet-epitaxy InAs/InP QDs have been developed to improve emitter symmetry and reduce fine-structure splitting, which is particularly important for entangled-pair generation. MOVPE-grown structures have achieved very low multi-photon emission \(g^{(2)}(0)=0.003\), although the raw HOM visibility remains limited to 19.3\% without post-selection\cite{HolewaSemenovaNatComm24}. Integration of such emitters into photonic crystal cavities has yielded Purcell factors around 5 and lifetimes down to 340\,ps\,\cite{PhillipsSciRep2024}. Overall, these results show that InAs/InP QDs are rapidly approaching the low multi-photon contribution and extraction performance required for telecom quantum networking, while indistinguishability and device-to-device uniformity remain the key metrics that still lag behind the best near-infrared QD sources.
    
For InAs/GaAs QDs, emission is shifted into the telecom range using strain-reducing layers (primarily O-band) or InGaAs metamorphic buffers (MMB) (O- and C-bands). Low multi-photon contribution C-band emission has been demonstrated with $g^{(2)}(0)=0.003$ from MOVPE-grown QDs on InGaAs MMB with AlAs/GaAs DBRs\,\cite{PaulMichlerAPL17}. Recent MMB optimizations have improved crystal quality and enabled integration into micropillar and circular Bragg grating cavities\cite{SittigNanoph23,SorokinJEPT25}, with ongoing progress in improving crystal quality and enhancing indistinguishability.

Comprehensive performance comparisons of these III-V telecom QD platforms for long-distance quantum networks are provided in recent reviews\cite{MichlerWaksLiuNatNano23,HolewaNanoPh25} and a recent review on telecom QDs in cavity devices can be found in Ref.\cite{HuberLoyolaAPR2026}. 
A comparison of state-of-the-art performance for near-infrared and telecom quantum-dot single-photon sources is presented in Tab.~\ref{tab:telecom-nir-performance}. While near-unity indistinguishability is routinely achieved in the NIR, telecom devices still lag in this key metric that is essential for quantum networking. Nevertheless, the rapid recent progress reported for C-band sources~\cite{KimHoeflingAQT2025} suggests that high-quality telecom QDs suitable for practical repeater applications may become available on short timescales.

\textbf{Entangled pair-sources\,(EPS)}
Near-infrared (NIR) entangled photon-pair sources (EPS) based on GaAs or InGaAs quantum dots have now reached near-unity entanglement fidelity. Record performance has been demonstrated with 69\% extraction efficiency, 0.96 fidelity, and 0.47 pair efficiency per pulse using circular Bragg gratings~\cite{RotaElight2024}. Cavity-integrated GaAs devices further confirm that material quality, rather than the biexciton--exciton cascade itself, is no longer the dominant limitation~\cite{HuberNatComm17}. Integration into bullseye cavities has recently revealed wavevector-dependent polarization mixing. This effect originates from the angle-dependent coupling between the emitter and the cavity mode~\cite{LaneveNatComm2025}. 

Telecom-band entangled photon-pair sources (EPS) have recently approached the performance levels of their near-infrared counterparts. 
In the O-band, site-controlled nanowire QDs have enabled on-demand polarization-entangled photon pairs with high fidelity at 
85.8\%\cite{AlqedraNanoLett25}. In the C-band, InAs/GaAs QDs on MMBs have produced on-demand entangled pairs with a peak fidelity of 96.4\% 
with measured pair rates of 200\,kcps \cite{JoosAQT26}. However, telecom EPS still lag behind NIR devices in brightness, 
indistinguishability, and entangled photon pair rates. These results underscore that while telecom EPS are now viable, further improvements 
in crystal quality and cavity design are required to match NIR performance for practical quantum repeater applications. 
 
%
%
%
\subsection{Protocols}
\label{Protocols}
\subsubsection{Extended spin coherence} 
The spin coherence time, $T_2$, can be significantly extended using spin Hahn echo and other dynamical decoupling techniques, such as Carr-Purcell (CP) and Carr-Purcell-Meiboom-Gill (CPMP) sequences. An overview of experimentally reported $T_2$ values in various QD systems is presented in figure \ref{DDQD}. Notably, coherence time exceeding 100~$\mu$s have been demonstrated recently in nano-hole droplet etched QDs, using CPMG pulse sequences \cite{ZaporskiLeGallNatNano23}. This improvement is attributed to enhanced strain uniformity and suppression of nuclear spin noise, resulted from better structure homogeneity during growth process \cite{NguyenWarburtonPRL2023, ShoferPRX2025, Koong2025}.\\
\begin{figure}[b!]
        \centering
        \includegraphics[width=0.99\linewidth]{Coherence_New.png}
        \caption{Summary of reported values for coherence time, $T_2$, in different QD systems. These results extracted from spin Hahn echo (square symbol) and other dynamical decoupling techniques, such as CP (triangle symbol) and CPMG (circle symbol). Blue and red colored symbols show the results from electron and hole spin coherence time. Shaded rectangular and circular regions show the results from nano-hole droplet etched QDs and QDM, respectively.}
        \label{DDQD}
\end{figure} 
Majority of these studies have focused on electron spin, which generally demonstrated longer spin coherence time \cite{GreilichScience2006, PressNatPhoton2010, VarwigPRB2013, StockillNatCommun2016, BodeyNPJQI2019, GangloffAtatureScience19, GangloffNatPhys2021, JacksonNatPhys2021, JacksonPRX2022, EversPSSB2024}; however, recently some studies explored the performance of hole spin in QDs \cite{DeGreveNatPhys2011, VarwigPRB2013, HuthmacherPRB2018, HoggNatPhys2025}. In addition, Quantum Dot Molecule\,(QDM) recently has shown spin coherence time on the order of hundreds of nanoseconds \cite{TranPRL2022}. For memory-based QR schemes where coherence time is a key parameter, these values indicate the growing potential of using QDs as memory storage nodes.
%
%
\subsubsection{Indistinguishability of remote QDs}
\label{SecIndistinguishability}
High indistinguishability between photons emitted by spatially separated quantum dots is essential for quantum repeaters. It requires precise spectral matching, active stabilization, and coherence preservation across independent devices and transmission channels. \\
%
Recent advances have demonstrated indistinguishability above $0.9$ for remote QDs. 
Resonantly driven GaAs quantum dots embedded in diode structures have achieved indistinguishability as high as 0.93 between remote emitters, enabled by active tuning and improved material quality \cite{ZhaiWarburtonNatNano22}. In the same work, a photonic CNOT gate was realized with a fidelity of 0.85, highlighting the relevance of high indistinguishability for quantum logic operations. Furthermore, long-distance interference experiments using frequency-converted photons have demonstrated non-classical interference over fiber links exceeding 300\,km, with raw visibilities of $V=0.67$ \cite{YouWeiPanAdvPhoton22}.

When generating entanglement through the biexciton-exciton cascade, as the finite lifetime of the biexciton reduces the quantum state purity which leads to a lower indistinguishability\cite{HuberOE2013, SchollPRL2020}. This is also limiting swapping fidelities\cite{BeccaceciArxiv2025} and other metric related to quantum repeaters. 
To remedy this, the biexciton has to be selectively Purcell enhanced, which recently has been demonstrated\cite{BehrendsArxiv2026}.
These results illustrate both the progress and remaining challenges. While near-unity indistinguishability is achievable for individual devices under controlled conditions, maintaining high visibility between remote sources over long distances remains limited by spectral mismatch, environmental noise, and conversion-induced imperfections. In the context of quantum repeaters, where two-photon interference underpins entanglement swapping, further improvements in source uniformity and frequency stabilization are essential.

%
    
\subsubsection{Gates, Entanglement and Teleportation}
\label{Gates}
Beyond high-quality photon sources, quantum repeaters also require robust local quantum processing at each node. Deterministic spin-photon gates, spin-photon entanglement, and quantum teleportation are essential for entanglement swapping, purification, and the coherent transfer of quantum information between photonic flying qubits and stationary spin qubits.

Coherent optical control of quantum dot spins on picosecond timescales is now well established. Waveguide quantum electrodynamics 
has enabled few-photon nonlinear phase shifts \cite{StaunstrupNatComm24} and inspired proposals for passive controlled-Z gates in 
chiral waveguide QED \cite{LevyYeyatiPRXQuantum25}. However, realizing deterministic high-fidelity spin-photon and photon-photon gates with sufficiently low error rates for practical quantum repeater operation remains an outstanding challenge.

Spin-photon entanglement was first realized in quantum dots over a decade ago \cite{GaoImamogluNature12, YamamotoDeGreveNature12, DelteilImamogluNatPhys15}. More recent on-chip demonstrations using time-bin encoding have reached fidelities of 0.7 with 86\% coupling efficiency \cite{ChanLodahlNPJQuantInf23}, making this a viable resource for both memory-based and all-photonic repeater architectures.
Spin-photon entanglement is also central to all-photonic quantum repeater architectures, where the quantum dot spin serves as a short-lived entangler to generate multi-photon linear cluster states and graph states\,(discussed in detail in Section~\ref{Memoryless}).


Quantum state transfer from a single photon to a distant quantum dot spin has been demonstrated using micropillar cavities in a Voigt geometry \cite{HePRL17}. More recently, quantum teleportation of a telecom-wavelength photon onto a GaAs quantum dot spin was achieved with a fidelity of 0.7 after frequency conversion \cite{StrobelNatCom25}.

Despite this progress, challenges remain in achieving higher gate and teleportation fidelities, better integration of spin-photon interfaces with photonic circuits, and compatibility with the long spin coherence times required for practical repeater operation.

In summary, 
III--V semiconductor QDs constitute a powerful and technologically mature platform for non-classical light generation in quantum repeaters. Their key strengths---near-unity photon indistinguishability, extremely low multi-photon emission probability, and deterministic on-demand operation---make them advantageous compared to probabilistic sources.
Two central challenges remain for practical QD-based quantum repeaters: (i) bringing the performance of telecom-wavelength QDs on par with the best near-infrared devices, and (ii) further extending spin coherence times, which would allow QDs to serve as a solid-state platform for photon generation and processing.    
    
\section{Repeaters}
\label{sec:Repeaters}
In the following section we discuss the two principal quantum repeater architectures. We first review all-photonic schemes based on QD-generated cluster and graph states in Section \ref{Memoryless}. We then turn to memory-based repeaters in Section \ref{Memory}, providing a brief overview of different physical platforms before focusing on the particularly promising heterogeneous architecture that combines deterministic QD photon sources with room-temperature alkali-vapor quantum memories. Within this context, we examine storage protocols and the current experimental progress toward their integration with QD sources. Quantum repeaters can also be classified according to how they handle loss and operational errors, which distinguishes between first-, second-, and third-generation schemes. This classification framework is primarily relevant to memory-based repeaters and will be discussed in Section \ref{Memory}.
%
%
%
%
\subsection{All photonic quantum repeater}
\label{Memoryless}
In 2015, Azuma \textit{et al.}~\cite{AzumaNatCom2015} proposed the concept of an all-photonic quantum repeater, motivated by the desire to eliminate one of the major technological bottlenecks of conventional repeater architectures: long-lived, high-performance quantum memories. Rather than generating entanglement between distant matter nodes and synchronizing probabilistic entanglement swapping through heralding and memory storage, the protocol relies entirely on large multipartite photonic resource states that distribute entanglement directly across the network. This architecture removes the need for matter-based quantum memories and long-distance two-way classical communication between repeater nodes, potentially enabling substantially higher communication rates. The central experimental challenge, however, is the generation, manipulation, and distribution of sufficiently large photonic graph states with high fidelity and low loss.
The main idea of their proposal is to change the order of information flow to swap the entanglement between the different parties. The Bell-state measurement, which is needed for the entanglement swap, can be seen as a controlled-Z gate, followed by an X-basis measurement of both photons. 
Instead of generating pairwise entangled photons, performing Bell-state measurements, and communicating the results classically, the protocol takes a different approach. It performs controlled-$Z$ gates directly between the photons that are sent out and carries out only the $X$-basis measurements locally at the locations where the Bell-state measurements would otherwise occur. 
To connect multiple parties there are still Bell- state measurements needed for swapping the entanglement. However, their approach has two advantages: first, there is no classical communication between the parties needed. The Bell-state result only has to be know locally and second, the protocol works completely without matter based quantum memory. 
A direct implementation this protocol starting with single photons would require effective controlled-Z interactions between photonic qubits, which has not been demonstrated 
yet. However, a controlled Z-gate between two qubits generates a graph-state~\cite{BriegelPRL2001,NielsenPRL2004}, so if one was 
able to generate a photonic graph-state, the proposal from Azuma et al. to generate a memory-less QR seems feasible. While the original proposal required graph states containing more than $10^6$ photons to reach communication distances of 1000\,km, subsequent protocol optimizations have dramatically reduced the required photonic resource overhead to the few-hundred-photon regime~\cite{BorregaardPRX2020}, with more recent one-way all-photonic architectures pushing this even further through more efficient error-correcting graph-state encodings~\cite{NiuNPJ2023,HilaireQuantum2021}. Zhan et al.~\cite{ZhanQuantum2023} recently published a performance analysis of different all photonic quantum repeaters based on matter spin enabled graph state generation. \\
While many different proposals exist to create more complicated graph states directly from a single 
emitter~\cite{ZhanPRL2020,ZhanQuantum2023,LiNpjQuantum2022,RussoPRB2018} or coupled single 
emitters~\cite{VezvaeePRAppl2022,GimenoSegoviaPRL2019}, experimental realizations have been limited to linear cluster-states, which are a 
special graph states, but could be used to fuse larger, more useful states~\cite{ThomasNature2024}. All experimental realizations, so far, for 
linear cluster state generation on quantum dots~\cite{SchwartzScience2016,CoganNatPhoton2023,CosteNatPhoton2023}, but also on neutral 
atoms~\cite{ThomasNature2022}, follow the proposal from Lindner and Rudolph~\cite{LindnerPRL2009}. Lindner and Rudolph suggested to use the 
electron or hole spin inside a quantum dot as a short lived local memory to entangle successively emitted photons in large chains, which is 
exactly the linear cluster-state that can be used as a building block for the above mentioned more complicated graph states. 
The main experimental challenge in implementing the Lindner–Rudolph proposal arises from its use of polarization encoding for the photons. Resonant excitation, which is required for high photon indistinguishability, typically demands polarization filtering to suppress scattered laser light, making polarization encoding difficult to realize. Usually, in QDs, polarization selection is used to suppress the 
reflection of the resonant laser~\cite{HeNatNano2013,KuhlmannRSI2013,GazzanoOptica2018,SchollNanoLett2019,WangNatPhoton2019}. Filtering a 
single polarization is obviously not compatible with polarization encoding. Thus, alternative methods have to be used. In the first 
demonstration of a linear cluster-state from solid state sources, Schwartz et al.~\cite{SchwartzScience2016} used the dark 
exciton~\cite{SchwartzPRX2015} to excited biexciton transition instead of a ground state charge spin. This is a very elegant solution, 
since the excitation laser excites the charge complex in an excited state and thus is not resonant to the collected photon. However, the 
indistinguishability of the emitted photon suffers from the added time jitter in the involved phonon transition and the fact that the dark 
exciton is not a stable ground state and has a finite energy. The authors were able to remedy the low indistinguishability of the photons 
by using a ground state hole~\cite{CoganNatPhoton2023}, as suggested in the original proposal from Lindner and Rudolph, and exciting the 
trion into an excited state, which is also a spin preserving excitation. Coste et al.\cite{CosteNatPhoton2023} used an electron instead of 
a hole, which has much shorter coherence time, and a LA photon assisted excitation scheme. However, their QD was embedded in a micropillar 
resonator and by exciting with a very high repetition rate, they were able to show a fidelity of $63\pm 5\%$ to a three photon linear 
cluster state, despite the short electron coherence time. Another proposal by Tiurev et al.~\cite{TiurevPRA2022} avoids the problem of 
polarization encoding by encoding in time-bin instead. Their proposal would also potentially allow for protocols to extend the spin 
coherence time~\cite{HuthmacherPRB2018,JacksonPRX2022,ZaporskiLeGallNatNano23}, however, it has not been demonstrated experimentally yet.
Despite all this great progress in cluster-state generation, the fusion of quantum repeater graph states or tree graph states is yet to be 
demonstrated. Furthermore, the manipulation of the states at the respective QR nodes will need fast detection and fast switching to allow for 
feed-forward, which likely has to be implemented in integrated optics, since bulk experiments with a few hundred photons do not sound 
feasible.

\textbf{Spin dephasing time} As discussed above, in all-photonic quantum repeater schemes based on quantum dots (QDs), the spin of a charged state often serves as an entangler for sequentially emitted single photons. While different photonic degrees of freedom, such as polarization and time-bin, can be used for encoding, the coherence of the entangler spin must be preserved throughout the operation. Therefore, the dephasing time $T^*_2$ of the spin becomes a critical figure of merit in evaluating the suitability of QDs for this protocol.

\begin{figure}[b!]
		\centering
		\includegraphics[width=0.99\linewidth]{DephasingTime_New.png}
		\caption{Experimentally reported dephasing time for single electron (blue markers) and hole (red markers) spins in QDs as a function of emission wavelength. Squares donate data from Ramsey experiments, whereas the circles show data from Ramsey experiments with additional spin cooling/locking techniques. Rectangles and circles  indicates results from nanohole droplet etched QDs and QDM, respectively.}
		\label{RamseyQD}
	\end{figure}
    
$T^*_2$ characterizes the timescale over which the phase coherence of the spin is preserved in the presence of environmental fluctuations, primarily due to the nuclear spin bath and nearby charge states. A summary of experimentally reported $T^*_2$ values is shown in Fig.~\ref{RamseyQD}. Notably, recent advances in nano-hole droplet-etched QDs have led to significant improvements in $T^*_2$, with values approaching 1$\mu$s \cite{NguyenWarburtonPRL2023, ShoferPRX2025, AppelNatPhys2025}. This improvement is attributed to enhanced of strain uniformity, which in turn reduced nuclear spin noise.

This figure includes data obtained from Ramsey experiments, both standard \cite{GreilichScience2006, BerezovskyAwschalomScience08, PressYamamotoNature08, PressNatPhoton2010, StockillNatCommun2016, StockillPRL2017, GangloffAtatureScience19, BodeyNPJQI2019, GangloffNatPhys2021, AppelPRL2021, JacksonNatPhys2021, LaccotripesNatCommun2024} and those incorporating advanced spin cooling/locking techniques \cite{JacksonPRX2022, NguyenWarburtonPRL2023, AppelNatPhys2025, ShoferPRX2025}.  These techniques are designed to suppress inhomogeneous broadening. Studies investigating hole spins are also included \cite{DeGreveNatPhys2011, GreilichNatPhoton2011, GoddenPRL2012, VarwigPRB2013, HuthmacherPRB2018, DusanowskiNatCommun2022, EversPSSB2024, Cizauskas2025a, CizauskasPRB2025b, Michl2025}. A noteworthy system to highlight is the Quantum Dot Molecule (QDM), a structure composed of two closely stacked QDs. A recent study reported a $T^*_2$ of approximately 100ns in a QDM system \cite{TranPRL2022}. Overall, these results reflect the significant progress in QD materials and device engineering, which has improved spin dephasing times in QDs.\\

    
\subsection{Memory-based quantum repeaters}
\label{Memory}
%
%
\subsubsection{Memory platforms}
\label{SteppingStonesMemoryTimes}
Different material platforms offer varying trade-offs in memory storage time, bandwidth, and operating conditions. 

Solid-state systems based on color centers and rare-earth ions have demonstrated some of the longest coherence times. 
The NV center in diamond combines electron spin coherence times of seconds with nuclear-spin registers enabling storage up to $\sim$60\,s \cite{BradleyPRX19}. 
SiV centers offer millisecond-scale coherence, $\sim$70\% indistinguishability, and $\sim$85\% emission efficiency \cite{BhaskarNature20}, with recent demonstrations of remote entanglement storage up to 500\,ms in a metropolitan network \cite{KnautLukinNature24}.
\begin{figure}[b!]
        \centering
        \includegraphics[width=0.99\linewidth]{StorageTime_NoTelecom_New_1.png}
       \caption{Reported storage times for different platforms used in memory-based quantum repeaters. 
       Data points are categorized by operational wavelength: black circles\,({Visible}) and black squares\,({NIR}). 
       The NIR category includes platforms operating in the near-infrared, such as single Rb atoms in cavities, cold atomic 
       ensembles, Rb BECs, and warm alkali vapors. Each data point is annotated with the corresponding reference.}
       \label{StorageTime}
\end{figure}

Rare-earth ions provide very good storage properties at cryogenic temperatures. In $^{151}$Eu$^{3+}$:Y$_2$SiO$_5$, spin coherence times reach $\sim$100\,ms and can be extended to hours using dynamical decoupling \cite{ZhongSellarsNature15}. At the single-photon level, storage times up to 1\,s have been achieved \cite{HainNJP25}. High-efficiency storage has also been demonstrated using atomic frequency comb protocols, with efficiencies exceeding 80\% and storage times up to 100\,ms \cite{JobezPRL16, OrtuNPJ22}. Telecom-compatible implementations have been realized in Er-doped crystals and fibers \cite{LagoRiveraRiedmattenNature21, SaglamyurekTittelNatPhoton15, JiangNatCom23}.

Cold atomic systems offer excellent performance in controlled environments. Storage times of up to 16\,s have been reported in 3D magneto-optical traps \cite{DudinKuzmichPRA13}, while single trapped atoms and ions have reached coherence times of 100\,ms and beyond one hour, respectively \cite{KorberRempeNatPhoton18, WangNatComm21}. In contrast, room-temperature alkali-vapor memories offer operational simplicity. 
Using EIT in buffered rubidium vapor, storage times up to 1.2\,ms have been demonstrated \cite{DujicScRep24}.

Figure~\ref{StorageTime} summarizes reported storage times across leading platforms, highlighting the wide range of achievable coherence times depending on the system and operating conditions.

Significant progress has been made in the last five years toward generating entanglement between remote quantum memories across several 
platforms. In diamond color centers, metropolitan-scale heralded entanglement was 
achieved between NV centers over 25\,km of deployed fiber~\cite{StolkHansonScAdv24}, while two nanophotonic SiV quantum memory nodes were 
entangled through a deployed telecom network including a 35\,km urban fiber loop with quantum frequency conversion~\cite{KnautLukinNature24}. 
In rare-earth doped crystals, heralded entanglement between remote Pr-doped memories was demonstrated at high rates with 
microsecond-scale storage~\cite{LagoRiveraRiedmattenNature21}. 
In atomic ensembles, heralded entanglement between remote memories was realized over 420\,km using a DLCZ-type protocol combined with quantum frequency conversion to the telecom S-band, exceeding the direct transmission limit~\cite{LuoPanarXiv25}. 
These results highlight the rapid development of memory-based quantum network links across different material platforms.


%
%
\subsubsection{Storage protocols}
\label{MemoryStorageProtocols}
The main storage protocols employed in ensemble-based quantum memories are Electromagnetically Induced Transparency 
(EIT), using a transparency window in a \(\Lambda\)-system; Off-Resonant Cascade Absorption (ORCA), a ladder-type 
protocol based on two-photon absorption; Controlled Reversible Inhomogeneous Broadening (CRIB), which reversibly 
broadens the atomic absorption line for storage and retrieval; and Atomic Frequency Combs (AFC), which uses a periodic spectral comb of absorption peaks to enable multimode storage. 

For interfacing quantum memories with broadband single-photon sources such as semiconductor quantum dots, 
the acceptance bandwidth of the memory becomes a key figure of merit, together with storage time, efficiency, 
and fidelity. These aspects will be reviewed below for the relevant protocols.

%

\textbf{Electromagnetically Induced Transparency\,(EIT)} offers considerable flexibility, since its performance can be tuned through the control-field strength, optical depth, and choice of atomic environment. This tunability enables optimization across different regimes of storage time, bandwidth, and efficiency. In cold atomic ensembles, EIT has achieved exceptionally long storage times of up to 16\,s with high fidelity~\cite{DudinKuzmichPRA13}. In warm vapors, storage times range from tens of microseconds~\cite{KupchakScRep15} to beyond 1\,ms in buffer-gas cells~\cite{DujicScRep24}, although longer storage generally comes at the cost of reduced bandwidth. Conversely, stronger control fields or optimized geometries can extend the acceptance bandwidth into the hundreds of MHz, making EIT compatible with broadband sources while preserving good coherence~\cite{WoltersPRL17}.

A practical limitation of EIT in alkali vapors, however, is the absence of suitable storage transitions at telecommunication wavelengths. This challenge has been addressed by combining EIT-based storage with frequency conversion. For example, in cold \(^{87}\)Rb atoms held in an extended dark magneto-optical trap, photons generated via spontaneous four-wave mixing were stored using EIT for 0.1\,s with 1.9\% efficiency, followed by frequency conversion with 55\% efficiency~\cite{KuzmichKennedyNatPhys10}. 
Comprehensive review of EIT-based quantum memories can be found in Ref.~\cite{NovikovaLaser&Photon12}.

%
    \begin{table*}
    \label{tab:TabII}
    \centering
    \small
    \begin{tabular}{|l|c|c|c|c|c|c|c|}
     \hline
    \textbf{Platform}  & \textbf{Protocol}         & \textbf{Fidelity}       & \textbf{T$_{\rm store}$} & \textbf{Efficiency}          & \textbf{SNR}         & \textbf{$\lambda$ (nm)} & \textbf{Bandwidth} \\
    \hline
    Rb\cite{WoltersPRL17} & EIT & -- & 50\,ns & 17\% & -- & 780 & 0.66\,GHz \\
    Rb (buffer gas) \cite{DujicScRep24}           & EIT               & --             & 1.2\,ms         & 25\%                & --           & 795            & kHz  \\
    Rb \cite{GuoPRL25}                           & Raman             & 98\%     & 20\,ns       & 82\%          & --           & 795            & 77\,MHz       \\
    Cs \cite{DavidsonCommPhys23}                        & ORCA/FLAME        & --             & 108\,ns         & 35\%                & noise-free   & 852            & \(\sim\)1\,GHz \\
    Rb \cite{HosseiniBuchlerNaturePhysics11}      & \(\Lambda\)-GEM   & 98\%           & 3\,\(\mu\)s     & 78\%                & --           & 795            & 500\,kHz  \\
    Rb\cite{SchofieldPhysRevLett26}             & AFC                 & 30\%           & 7.5\,ns         & 6.59\%    &   16.1 & 780 & 150\,MHz \\
    Rb (EDMOT) \cite{KuzmichKennedyNatPhys10}     & EIT/QC            & --             & 0.1\,s          & 1.9\% / 55\%        & --           & 793/1300       & MHz       \\
    \hline
     \hline
    \textbf{Cs\,(ladder)} \cite{MaaSSQST25} & (ORCA-like) & -- & 19.8\,ns & 0.6\%  & -- & 895 & $\sim$5\,GHz \\
      \textbf{Rb (ORCA)} \cite{ThomasWalmsleyPortalupiScAdv24} & ORCA & -- & 800\,ps & 12.9\% & 18.2 & 1529 & $\sim$GHz \\  
    \hline
  \end{tabular}
  \caption{Performance of alkali-based quantum memory protocols, predominately warm vapors with only \cite{KuzmichKennedyNatPhys10} using cold ensemble. Bandwidth is a critical parameter for compatibility with 
  deterministic QD single-photon sources (bottom two rows).}
  \label{tab2}
  \end{table*}

\textbf{Off-resonant Raman interactions} in warm atomic vapors have emerged as one of the leading protocols for broadband quantum storage. The first single-photon broadband Raman memory with 30\% efficiency and 4\,\(\mu\)s storage time was demonstrated in Rb vapor \cite{ReimWalmsleyPRL11}. This was quickly followed by the first storage of GHz-bandwidth non-classical photons in Cs Raman memories \cite{MichelbergerWalmsleyNJP15}. 
Raman protocols offer significant advantages through their large acceptance windows, routinely reaching the GHz range.
Subsequent advances using optimal control and improved cell preparation have pushed efficiencies above 82\% with unconditional fidelities reaching 98\% at the single-photon level \cite{GuonatCom19}, and very recent work has demonstrated near-unity efficiency and fidelity in warm Rb vapor but with compromised storage time\cite{GuoPRL25}. 

%
%

\textbf{Off-Resonant Cascade Absorption (ORCA)} has become one of the leading protocols for high-bandwidth, noise-free quantum storage in 
warm atomic vapors. It routinely achieves acceptance windows of \(\sim\)1\,GHz, making it particularly attractive for interfacing with 
broadband sources such as quantum dots. The original demonstration reached 16.7\% efficiency with a signal-to-noise ratio of 
$\propto 10^{-5}$ (designated "noise-free") at a storage time of a few nanoseconds \cite{KaczmarekNunnWalmsleyPRA18}. Subsequent improvements using 
dressing fields and cavity enhancement have pushed end-to-end efficiencies to 35\% while maintaining noise-free operation 
\cite{DavidsonCommPhys23, SrivathsanPRL25}. The short storage times, currently limited to \(\sim\)100\,ns by Doppler broadening and 
motion-induced dephasing, remain the main limitation.
%

\textbf{Controlled Reversible Inhomogeneous Broadening (CRIB) and its Gradient-Echo variant (GEM)} have delivered some of the highest single-mode efficiencies among quantum memory protocols. In warm alkali vapors, GEM has reached efficiencies up to 87\% \cite{HosseiniNatPhys10}. Extended \(\Lambda\)-GEM protocols in warm Rb vapor have demonstrated 98\% fidelity, 3\,\(\mu\)s storage time, and 78\% efficiency with tunable bandwidth \cite{HosseiniLamNatComm10}. In cold atomic ensembles, GEM has achieved \(\sim\)80\% efficiency with storage times up to 195\,\(\mu\)s \cite{SparkesNJP13}. 

%

The \textbf{Atomic Frequency Comb (AFC)} protocol, traditionally implemented in rare-earth-doped crystals at cryogenic temperatures\cite{JobezPRL16,OrtuNPJ22,ZhongNature15,TittelQST2025}, has recently been demonstrated in room-temperature alkali vapors. For the first time, storage at a single photon level had been reported\cite{SchofieldPhysRevLett26} in warm $^{87}$Rb vapor, achieving a storage time of 7.5 ns with an efficiency of 6.59\% and a signal-to-background ratio of 16.1. 
While these results represent an important progress, AFC storage in alkali vapors remains at an early stage and currently offers lower performance compared to EIT and Raman protocols.


Among the protocols discussed above, ORCA and Raman are particularly well suited for interfacing with quantum-dot sources owing to their large acceptance bandwidths ($\sim$1 GHz). EIT offers complementary advantages through its tunability of bandwidth and efficiency via optical depth and control-field strength. These features make the protocols discussed here good candidates for the heterogeneous QD--alkali-vapor interfaces examined in the following section.
For comparison between different protocols refer to Tab.\ref{tab2}. %

%
\subsubsection{Heterogeneous QD–Alkali-Vapor Interfaces}
\label{HybridQDs}
%
While most quantum storage experiments have relied on weak coherent pulses or probabilistic sources, practical quantum 
repeaters require deterministic single-photon sources -- such as semiconductor QDs -- that substantially improve entanglement distribution rates, see Section \ref{benchR}. 

Experimental demonstrations of QD–alkali interfaces have progressed rapidly. Early works showed resonance fluorescence from quantum dots 
interfaced with cesium vapor~\cite{MichlerPRB14}, and, more recent, investigation of retardation effects in Cs 
vapor~\cite{SchmidtBensonScRe19}. 

Rayleigh regime in resonant fluorescence is particularly promising due to its narrow spectral bandwidth. 
Therein, the coherent\,(elastic) scattering component dominates and inherits the narrow linewidth of the excitation 
laser\,(instrument-limited to ~27 MHz\cite{MaleinPRL16} as shown at NIR). 
Resonance fluorescence has also been demonstrated for O-band QDs \cite{Al-KhuzheyriGerardotAPL16} and, 
more recently, for C-band QDs \cite{WellsShieldsNatComm23}.
However, some studies suggest that the narrow-band coherent scattering component may lose its single-photon character under 
strong spectral filtering\cite{HanschkePRL20}. The suitability of this approach for quantum networking applications therefore 
still requires experimental verification.

Storage of single photons from semiconductor quantum dots has been demonstrated using telecom-wavelength ORCA memories, achieving deterministic storage and retrieval with 13\% efficiency \cite{ThomasWalmsleyPortalupiScAdv24}. However, practical use remains limited by short storage times, currently in the sub-nanoseconds range. In parallel, four-wave mixing in rubidium vapor has been proposed as a route toward telecom-compatible interfaces with QDs\,\cite{ZajacWashington25}. More recently, storage of QD photons in a Cs vapor ladder-type memory for up to 20\,ns was demonstrated\cite{MaaSSQST25}. 

For these interfaces to deliver competitive performance in quantum repeaters, further improvements in memory efficiency $\eta_m$ and coherence time $\tau_m$ are essential. As analyzed in Sec.\ref{benchR}, even moderate gains in these parameters can substantially increase entanglement distribution rates by mitigating the exponential decay of memory efficiency during the waiting time between successful entanglement generation and swapping. Continued progress in cavity-enhanced storage protocols, optical-depth optimization in EIT, and improved spectral matching between QD emission and memory acceptance windows should enable heterogeneous QD--alkali-vapor systems to advance from current proof-of-principle experiments toward practical first-generation repeaters.
%
%
%
%
\subsubsection{Entanglement distribution rates}
\label{benchR}
Memory-based quantum repeaters are commonly classified into three generations according to how they suppress loss and operational errors~\cite{JiangLukinScRep16}. In first-generation schemes, both loss and operational errors are suppressed probabilistically via heralding. In second-generation schemes, loss errors remain probabilistic while operational errors are corrected deterministically using quantum error correction. In third-generation schemes, both types of errors are suppressed deterministically.

To compare the resource efficiency of these generations, Jiang et al. introduced a cost function, with a more practical metric being the cost coefficient $C'(L_{\rm tot})$, which quantifies the overhead (qubits $\times$ time) required to generate one secret bit per kilometer. This coefficient depends on key hardware parameters such as gate error probability $\epsilon_G$, coupling efficiency $\eta_c$, and local operation time $t_0$.
Analysis of $C'(L_{\rm tot})$ reveals distinct operating regimes: first-generation schemes perform best for high gate errors ($\epsilon_G \gtrsim 1\,\%$). Second-generation schemes with encoding become advantageous for intermediate gate errors ($\epsilon_G < 1\,\%$) combined with poor coupling efficiency ($\eta_c < 90\,\%$) or slow operations ($t_0 > 1\,\mu$s). Third-generation architectures outperform the others only when gate errors are low ($\epsilon_G < 1\,\%$), coupling efficiencies are high ($\eta_c > 90\,\%$), and operations are fast ($t_0 < 1\,\mu$s).

Most current experimental demonstrations, including heterogeneous systems based on quantum dots and room-temperature alkali-vapor memories, still operate within the first-generation regime. 

Building on the cost coefficient introduced before, we now perform a hardware-specific rate analysis focused on near-term memory-based architectures. For that, we use entanglement distribution rate \( R \) (in $\mathrm{s^{-1}}$), which quantifies the average number of successfully distributed high-fidelity entangled pairs per second over the total link distance. \( R \) depends on hardware and protocol parameters, including memory efficiency \( \eta_m \) and coherence time \( \tau_m \), detector efficiency \( \eta_d \), fiber attenuation \( \alpha_\lambda \) at the transmission wavelength, source efficiency \( \eta_S \), as well as the specific implementation of the Bell-state measurement via probabilities of successful entanglement per link \( p_0 \) (single-photon\,(1+1) or two-photon\,(2+2) detection, with or without post-selection).

\begin{figure}[b!]
        \centering
        \includegraphics[width=0.99\linewidth]{Entanglement_rate_Originlab_fifth.png}
        \caption{Entanglement distribution rate versus total distance for heterogeneous QD--alkali-vapor quantum repeaters using the model of Ref.~\cite{WuPRA20}. Solid lines correspond to the 1+1 BSM protocol and dashed lines to the 2+2 protocol. 
        Three scenarios with different memory lifetimes, source efficiencies, and memory efficiencies are compared:
        \textcolor{blue}{Blue}: $\tau_m = 1$ s, $\eta_s = 0.15$, $\eta_m = 0.7$;
        \textcolor{red}{Red}: $\tau_m = 1.2$ ms, $\eta_s = 0.9$, $\eta_{spp} = 0.7$, $\eta_m = 0.15$;
        \textcolor{Green}{Green}: $\tau_m = 1.2$ ms, $\eta_s = 0.9$, $\eta_{spp} = 0.7$, $\eta_m = 0.7$.}
        \label{fig:dlczVSdirect}
   \end{figure}
Sangouard \textit{et al.} \cite{SangouardGisinRevModPhys11} compared the entanglement distribution time ($1/R$) as 
a function of distance for atomic-ensemble-based schemes as a function of nesting levels $n=0,1,....,4$. The basic 
DLCZ protocol \cite{DuanLukinCiracZollerNature01} served as the slowest baseline, while the single-photon source 
approach \cite{SangouardPRA07} boosted the rate by one to two orders of magnitude.
This improvement arises from replacing the original DLCZ's probabilistic light sources with near on-demand single-
photon emitters. QD sources deliver exactly this benefit: their effective emission probability 
increases from $p \leq 0.01$ (the value typically used in DLCZ to suppress multi-photon errors) to $p \approx 1$ 
in an ideal case. 
However, that particular model implementation\cite{SangouardGisinRevModPhys11} assumes ideal memories with 
infinite coherence times what leads to significant overestimates of both the rates and the 
crossing point with direct transmission.

Another important rates improvement was proposed in Ref.~\cite{ChenWeiPanPRA07}, where the single-photon 
Mach--Zehnder interference of the original DLCZ protocol was replaced by two-photon Hong--Ou--Mandel (HOM) 
interference for entanglement generation and swapping. This relaxes the phase stability requirement from 
sub-wavelength ($\sim$0.1\,$\mu$m) control over km-scale links thereby enabling substantially higher entanglement 
distribution rates.

{Further improvements were achieved using two-photon Bell-state measurements. The photon-pair source combined with two-photon Bell-state measurement~\cite{SimonGisinPRL07} achieves about 2--3 orders of magnitude improvement over the original DLCZ protocol. Moreover, the use of multi-mode memories, such as rare-earth-ion-doped crystals with atomic frequency comb techniques, compare Section \ref{MemoryStorageProtocols}) leads to the highest rates, reaching roughly 3--4 orders of magnitude better performance than DLCZ at long distances.}
Overall, the advanced schemes provide up to 4 orders of magnitude higher entanglement distribution rates than the basic DLCZ protocol for this model\cite{SangouardPRA07}. 
%

Hardware-driven rate comparison focusing on near-term hardware parameters and realistic ensemble-based quantum memories was presented in Ref.\cite{WuPRA20}. 
{In their model, entanglement generation attempts in each elementary link occur probabilistically, while the memory efficiency decays exponentially in time according to $\eta_m(t) = \eta_m(0) \exp(-t / \tau_m)$. This explicit time dependence of the memory efficiency during the waiting time between successful link generations makes the swapping probability—and therefore the overall entanglement distribution rate—strongly dependent on memory lifetime, providing a significantly more realistic description than models assuming constant memory performance.}

Adapting their model, Fig.~\ref{fig:dlczVSdirect} illustrates the entanglement distribution rate versus total distance for 
heterogeneous QD--alkali-vapor quantum repeaters at nesting level $n=1$. 
{All curves are calculated using the model of Wu et al., with different memory coherence times: $\tau_m = 1$ s for a long memory 
storage time and low source-efficiency scenario and $\tau_m = 1.2$\,ms\cite{DujicScRep24} for a reported in literature 
memory storage time and QDs source-efficiency. We vary both the source efficiency 
$\eta_s$ (and entangled-pair efficiency $\eta_{spp}$ where applicable) and the effective memory efficiency $\eta_m$, which 
now was extended to incorporate bandwidth matching between the QD emission and the alkali-vapor memory.} 
For the modeling we use $\eta_m = 0.1$, consistent with recent experimental demonstrations of QD photon storage in telecom ORCA 
 memories~\cite{ThomasWalmsleyPortalupiScAdv24}. We further adopt $\eta_s = 0.9$ (and $\eta_{spp} = 0.7$ for entangled-pair sources), representing high but currently achievable efficiencies for QD sources.
 
 The plot shows that at short distances the green curves (high source efficiency combined with high memory efficiency) yield the highest 
 entanglement distribution rates. However, due to the relatively short memory lifetime of 1 ms, these rates decay rapidly with distance. In 
 contrast, the blue curves, which use a much longer memory coherence time of 1 s despite lower source efficiency, maintain significantly 
 higher rates beyond approximately 100 km, highlighting that memory lifetime becomes the dominant limiting factor at longer distances. 
 Finally, the red curve demonstrated the importance of memory coupling, $\eta_m$, for high performance repeaters schemes.
 
The model compares the {1+1} and {2+2} Bell-state measurement protocols. For the {1+1} protocol, post-selection is 
used to convert the heralded state into a usable two-photon entangled state suitable for quantum communication. 
In our case, the {1+1} scheme outperforms the {2+2} scheme when using a long memory lifetime of $\tau_m = 1$\,s. 
However, for a shorter memory lifetime of $\tau_m = 1$\,ms, the {2+2} scheme becomes advantageous, delivering rates 
{2--3 orders of magnitude higher}, depending on memory efficiency.

{Combining these achievable storage times, memory efficiencies\,(including bandwidth matching), and the high source efficiencies achievable with quantum dots, we provide a near-to-medium-term benchmark of entanglement distribution rates for heterogeneous QD--alkali-vapor systems.}
%
%
\section{Enabling technologies}
While Sections \ref{sec:QDs} and \ref{sec:Repeaters} established the strong potential of III–V quantum dots as deterministic photon sources and evaluated both all-photonic and memory-based repeater architectures, translating these advances into practical long-distance fiber networks requires addressing several engineering bottlenecks.
Optimal QD performance demands cryogenic temperatures to achieve the required spin coherence and photon indistinguishability. Network scaling additionally requires compact photonic integration, sub-nanosecond synchronization across nodes, and multiplexing to increase effective rates. Together, these are critical enabling technologies for deployable QD-based quantum repeaters.

In the following subsections we highlight their current maturity, challenges, and prospects, with emphasis on compatibility with QD sources and heterogeneous QD–alkali-vapor memory interfaces.
\label{technology}
\subsection{Cryo-coolers}
Atomic vapors offer a major practical advantage - they operate at room temperature unlike all other quantum memory platforms discussed in Sec.~\ref{SteppingStonesMemoryTimes}.
In contrast, III-V QDs exhibit optimal single-photon emission characteristics at cryogenic temperatures, 
typically around 4\,K. At elevated temperatures, phonon interactions cause significant broadening of the emission lines, increased 
dephasing, and degradation of multi-photon contributions and indistinguishability~\cite{SchlehahnReitzensteinRevScInstr15,ZajacPRB16}.

Achieving and maintaining these low temperatures has historically relied on complex and resource-intensive systems. Early experiments 
predominantly used bath liquid-helium cryostats, which require frequent manual refills. Modern closed-cycle cryocoolers have largely replaced them, yet many laboratory-scale systems remain bulky, power-hungry 
(often consuming several kilowatts), and ill-suited for deployment in real-world quantum network nodes.
For practical quantum communication infrastructures, single-photon sources must operate in intermediate network nodes that are 
robust, user-friendly, rack-mountable, and compatible with data-center environments. This requirement has driven the development of compact, low-vibration cryocoolers with reduced footprint and power consumption. Particularly promising are linear Stirling-type cryocoolers, which offer a favorable trade-off between size, efficiency, and base temperature~\cite{SchlehahnReitzensteinRevScInstr15,SchlehahnScRep18}.

Initial implementations faced challenges related to mechanical vibrations and higher base temperatures, which can degrade nano-photonic performance. However, these limitations have been substantially mitigated in recent years. A key demonstration employed a twin-piston linear Stirling cryocooler that reached a base temperature of approximately 29\,K (unloaded) with excellent temperature stability ($\Delta T \approx 0.05$\,K) and reduced vibration amplitude ($\sim$1\,$\mu$m at 46\,Hz). The system exhibited a power consumption of only 280\,W---roughly an order of magnitude lower than conventional closed-cycle systems---while producing intensity fluctuations reduced by a factor of three compared to He-flow cryostats (for a typical 3\,$\mu$m excitation spot)~\cite{SchlehahnReitzensteinRevScInstr15}.

Characterization of QD-based single-photon sources on this platform at $\sim$30\,K revealed linewidth broadening to $\sim$20\,$\mu$eV (versus a few $\mu$eV at 4\,K), increased $g^{(2)}(0)$ values around 0.2
, and reduced photon indistinguishability (visibility dropping from 98\% at 4\,K to 40\% at 30\,K). Polarization-entangled photon pair fidelity showed only modest degradation, from 72\% at 4\,K to 68\% at 30\,K~\cite{HafenbrakSchmidtNJPhys07,SchlehahnReitzensteinRevScInstr15}.

Subsequent engineering efforts have further advanced stand-alone, fiber-coupled QD sources integrated into compact Stirling cryocoolers, demonstrating plug-and-play and rack mounted operation~\cite{SchlehahnScRep18,MusialAQT20}. These developments significantly improve the prospects for field-deployable quantum light sources.
Nevertheless, the requirement for cryogenic operation remains a major obstacle to widespread deployment. Ongoing research therefore pursues both improved miniature cryocoolers (including advanced pulse-tube and hybrid designs) and material strategies that enable higher-temperature operation of high-performance QDs, aiming to bridge the gap toward practical, scalable quantum networks~\cite{BremerMQT22}.
%
%
\subsection{On-chip integration}
Heterogeneous integration of III-V QDs with photonic integrated circuits (PICs) can be achieved through three main approaches: monolithic fabrication, transfer printing, and wafer bonding.

High performance has been achieved using photonic-crystal waveguides (PCWs) in {monolithic GaAs platforms}. Near-unity emitter-to-waveguide coupling efficiencies ($\beta \approx 98\%$) were demonstrated~\cite{ArcariPRL14}, with tapered out-couplers enabling chip-to-fiber coupling efficiencies exceeding 80\%, resulting in overall source efficiencies above 10\% while preserving low multi-photon emission and high indistinguishability~\cite{DaveauOptica17}.

A versatile heterogeneous method is {transfer printing}, which enables high-precision, deterministic placement of pre-fabricated components. Telecom-band InAs/InP QD single-photon sources embedded in photonic crystal cavities have been transfer-printed onto CMOS-processed silicon waveguides~\cite{KatsumiAPEX23}. Transfer printing has also been used for micro-assembly of silicon photonic crystal cavity (PhCC) arrays, where individual cavities were measured \textit{in-situ}, sorted by resonance, and precisely placed to overcome fabrication variability while preserving high-Q suspended geometries~\cite{StrainNatComm25,StrainAPR22}.

{Wafer bonding} has also been successfully employed. Direct bonding of InAs/InP QDs on silicon-on-insulator (SOI) chips yielded an overall single-photon efficiency of $\sim$5\% in the telecom band~\cite{BurakowskiSyperekOptExpress24}. Similarly, heterogeneous integration of GaAs quantum dots with silicon nitride (Si$_3$N$_4$) platforms via wafer bonding and optimized mode matching achieved up to 98\% emitter-to-waveguide coupling~\cite{DavancoSrinivasanNatComm17}.

Inverse design techniques have enabled automatic optimization of complex photonic components (couplers, cavities, demultiplexers, and power splitters) with superior performance, compactness, and foundry compatibility. In particular, inverse-designed grating couplers have achieved excellent out-coupling efficiencies, including designs with peak efficiencies around 30\% in GaAs membranes~\cite{CarfagnoACSPhoton23}. Additional demonstrations include inverse-designed compact power splitters, wavelength- and mode-division multiplexers, and high-Q resonators that are suitable for integration with III-V quantum dot platforms.

For complete on-chip quantum functionality, efficient single-photon detection is essential. Waveguide-integrated superconducting nanowire single-photon detectors (SNSPDs) currently offer the best performance, with near-unity on-chip detection efficiency and low dark counts. Large-scale foundry-compatible integration of such detectors on silicon nitride platforms has already demonstrated a median on-chip detection efficiency of 93.4\% in complex circuits~\cite{PsiQuantumNat25}.

For comprehensive reviews on the on-chip integration of quantum emitters with photonic circuits, see Refs.~\cite{DietrichScience16,UppuLodahlNatureNano21,MoodyJPP22}.

The combination of high-efficiency QD integration, near-unity on-chip coupling, low-loss waveguides, inverse-designed components, and integrated high-performance SNSPDs is paving the way for fully integrated quantum photonic circuits suitable for practical quantum networks and optical quantum computing.
\subsection{Network synchronization}
Classical synchronization signals co-propagating with quantum signals in the same optical fiber remain a major source of noise due to their orders-of-magnitude higher intensity. The dominant crosstalk mechanisms are four-wave mixing and spontaneous Raman scattering, as previously characterized in quantum key distribution systems~\cite{PetersTyagiNJPhys09}.
The short coherence time of quantum dot single-photon sources, typically around 500\,ps, places severe constraints on acceptable timing jitter. Nanosecond-level synchronization, while useful in early demonstrations, is insufficient to maintain high photon indistinguishability and entanglement fidelity over deployed links.

GPS-based timing synchronization has been implemented in reconfigurable quantum local area networks (QLANs), achieving nanosecond jitter that was shown to limit remote state preparation fidelity~\cite{AlshowkanLukensPRX21}. That work also introduced the practically relevant metric of entangled bits per second (ebit/s) for entanglement distribution performance. 
More recent experiments demonstrated co-propagation of classical signals at 1300\,nm with 100\,GHz-bandwidth quantum channels at 1500\,nm over 100\,km using White Rabbit Precision Time Protocol (PTP), which offers nominal sub-nanosecond accuracy~\cite{BurenkovPolyakovOpticsExpress23,LipinskiAlvarezIEEE11}.

Nevertheless, reaching stable synchronization well below the 500\,ps coherence time of typical QD sources continues to be a key challenge for scalable quantum networks. Hybrid synchronization architectures combining White Rabbit with optical two-way time transfer or active phase stabilization will likely be required to meet the stringent demands of solid-state emitter-based quantum communication.

\subsection{Multiplexing}
\label{sec:multi}
Multiplexing constitutes one of the most effective routes to overcome the rate limitations shown in Fig.\ref{fig:dlczVSdirect} by 
parallelizing heralding, storage, and Bell-state measurements. 
Temporal/random-access and spatial multiplexing had bee shown in warm alkali vapors, in particular, 
up to four independent pulses have been stored in a single Cs cell with EIT (internal efficiency 36\%)\cite{MessnerOptExp23}, while nanoprinted on-chip light-cage arrays demonstrate scalable spatial multiplexing inside one room-temperature vapor cell\cite{GomezArxiv2025}. 
Early landmark experiments in cold atomic ensembles already achieved extreme spatial 
multiplexing with 225 individually addressable memory cells in a macroscopic ensemble using acousto-optic beam steering\cite{PuNatComm17}. 

REI crystals provide large inhomogeneous broadening, 
enabling temporal, spectral, and spatial multiplexing via the atomic frequency comb (AFC) protocol, 
with demonstrated capacities of tens to hundreds of modes\cite{OrtuNPJ22,TellerNPJQI25}. 
Optical frequency combs serve two key functions in this context. They enable the preparation of high-fidelity, broadband atomic 
frequency comb (AFC) structures through simultaneous multi-frequency spectral hole burning, and they support true frequency-multiplexed operation. The latter provides a natural interface to spectrally multiplexed quantum dot sources or frequency-converted photons while 
allowing the total number of modes to approach the $10^3$ regime required for near-deterministic repeater performance~\cite{TittelQST2025,MasukoApplOpt24,SinclairTittelPRL14}.
This multiplexing functionality makes both the QD–alkali hybrid and homogeneous REI platforms highly competitive for 
practical network application.

\section{Summary}
State-of-the-art III--V QDs deliver deterministic, high-rate single-photon and entangled-pair sources with record-low multi-photon emission probabilities and near-unity indistinguishability, both in the near-infrared and, increasingly, in the telecom O- and C-bands. 
We reviewed two complementary repeater architectures: all-photonic schemes that exploit the high repetition rates and on-demand character of QD sources, and memory-based repeaters that combine QDs as photon sources with alkali-vapor quantum memories. Particular emphasis was placed on heterogeneous architectures, where the deterministic emission of QDs is paired with the room-temperature operation and long coherence times of warm alkali ensembles. Compatibility between QD emission spectra and the main ensemble storage protocols (EIT, ORCA, CRIB, and AFC) was discussed in detail with entanglement rates estimated for near-term hardware.
Finally, we highlighted enabling technologies critical for real-world deployment listening both recent progress and remaining engineering challenges.
Taken together, these advances indicate that heterogeneous QD--alkali-vapor platforms constitute a highly scalable route toward practical, metro- and long-haul quantum networks. Importantly, the current limitations appear to be primarily technological rather than fundamental. Continued improvements in telecom QD performance, memory bandwidth, and system integration will therefore be decisive in realizing the full potential of quantum repeaters based on quantum dots.
     
\section{Acknowledgments}
JMZ acknowledges partial support from the Early Career Award\,(ECA) from the U.S. Department of Energy, Office of Science, Quantum Information Science and Technology (QIST) program. Part of this work was carried out while JMZ was employed at Brookhaven National Laboratory. JMZ is currently an independent researcher. 
THL acknowledges financial support from the German ministry of science and education (BMBF) within the Project Qecs (FKZ: 13N16272). THL and SH thank furthermore the BMBF for support within the project QR.X (FKZ: 16KISQ010) and QR.N (FKZ: 16KIS2209).

JMZ acknowledges helpful discussion with C.Simon and Y.Wu on the repeater modeling.

\bibliography{QDsreviewJZ.bib}	
\end{document}